# Scattering properties of PT-symmetric objects

**Mohammad-Ali Miri,[1,4,*] Mohammad Amin Eftekhar,[1] Margarida Facao,[2] Ayman F. Abouraddy,[1] Ahmed Bakry[3], Mir A. N. Razvi[3], Ahmed Alshahrie[3], Andrea Alù[4], and Demetrios N. Christodoulides[1,3]**

[1]CREOL, College of Optics and Photonics, University of Central Florida, Orlando, Florida 32816-2700, USA

[2]Department of Physics, I3N, University of Aveiro, Campus Universitário de Santiago, 3810-193 Aveiro, Portugal

[3]Physics Department, King Abdulaziz University, Jeddah 21589, Saudi Arabia

[4]Department of Electrical and Computer Engineering, The University of Texas at Austin, Austin, Texas 78712, USA

*Corresponding author: alimiri@utexas.edu*

**Abstract**

We investigate the scattering response of parity-time (PT) symmetric structures. We show that, due to the local flow of energy between gain and loss regions, such systems can deflect light in unusual ways, as a function of the gain/loss contrast. Such structures are highly anisotropic and their scattering patterns can drastically change as a function of the angle of incidence. In addition, we derive a modified optical theorem for PT-symmetric scattering systems, and discuss its ramifications.

**I. Introduction**

Non-Hermitian systems have recently attracted considerable attention in many and diverse areas of physics. This flurry of activity has been sparked within the framework of quantum field theory after recognizing that a wide class of non-Hermitian Hamiltonians that respect PT symmetry can exhibit entirely real spectra [1,2]. Lately, it was shown, both theoretically and experimentally, that PT-symmetric potentials can be realized in optics by judiciously incorporating gain and loss [3-4]. Since then, PT-symmetric optical structures have been intensely studied in a number of settings [5-26]. As it has been shown, the presence of balanced gain and loss regions in such structures can lead to a wide range of interesting phenomena such as negative refraction [10], unidirectional invisibility [11-13], pseudo-Hermitian Bloch oscillations [6,14] and single mode lasing [22-24], to mention a few. Thus far, most of these studies have been focused on guided wave systems and lattices. On the other hand, much less attention has been paid to scattering arrangements in two- and three-dimensional geometries [7,9]. Therefore, it is of interest to explore the effect of PT symmetry on the scattering response of electromagnetic waves. In this Letter, we study scattering of light from PT-symmetric dielectric objects. To elucidate this behavior, we consider an infinitely long Janus-like cylinder that involves both gain and loss in a fully symmetric fashion as shown in Fig.1. We show that such a structure can deflect scattered light by an amount that is related to its gain/loss contrast. In addition, as we will see, such objects are highly anisotropic, and the far-field scattering pattern can change with the angle of incidence. Finally, we discuss a modified optical theorem that is applicable to PT-symmetric structures.

## II. Mathematical formulation

In general, a dielectric object respects PT symmetry provided that its relative electric permittivity satisfies [2]

$$\epsilon^*(-\boldsymbol{r}) = \epsilon(\boldsymbol{r}). \quad (1)$$

This latter relation directly indicates that, for this symmetry to hold, the real part of permittivity (or refractive index) must be an even function of the position vector while its imaginary (gain/loss profile) must be antisymmetric. For example, this condition can be readily observed in homogeneous (in terms of their refractive index) Janus spherical or cylindrical configurations (Fig.1), where one half exhibits gain while the other an equal amount of absorption. Other more involved PT-symmetric patterns can also ensue from Eq.(1) in both 2D and 3D systems.

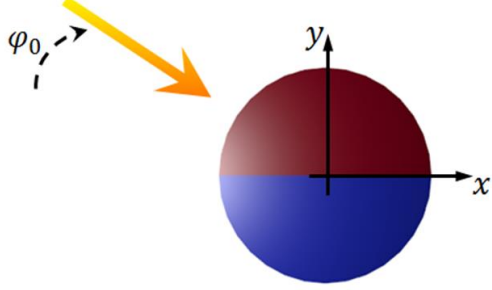

**Fig.1.** Plane wave incident on a PT-symmetric Janus cylinder

To demonstrate these effects, let us consider a two-dimensional dielectric body in the $xy$ plane. For reasons of simplicity, we restrict our analysis to the TE case where the electric field component $E_z = E(\mathbf{r})e^{-i\omega t}$ is perpendicular to the plane of propagation. In this case, the electric field obeys:

$$\nabla^2 E(\mathbf{r}) + k^2 \epsilon(\mathbf{r}) E(\mathbf{r}) = 0, \quad (2)$$

where in this notation, $\nabla = \hat{\mathbf{x}} \partial/\partial x + \hat{\mathbf{y}} \partial/\partial y$, $\mathbf{r} = \hat{\mathbf{x}} x + \hat{\mathbf{y}} y$, and $k = 2\pi/\lambda$ represents the wavenumber in the background medium (of permittivity $\epsilon_b$) and finally $\epsilon(\mathbf{r}) = \epsilon_m(\mathbf{r})/\epsilon_b$ corresponds to the normalized spatial distribution of the relative permittivity of this object $\epsilon_m(\mathbf{r})$, which is in general a complex quantity. When a dielectric object is illuminated by an arbitrary incoming wave, the total electric field can always be decomposed in terms of incident $E^{(i)}$ and scattered $E^{(s)}$ components as follows $E(\mathbf{r}) = E^{(i)}(\mathbf{r}) + E^{(s)}(\mathbf{r})$, where the incident field satisfies the Helmholtz equation in the background medium $\nabla^2 E^{(i)}(\mathbf{r}) + k^2 E^{(i)}(\mathbf{r}) = 0$. Therefore the scattered field should satisfy an integral equation $E(\mathbf{r}) = E^{(i)}(\mathbf{r}) + \frac{ik^2}{4} \int (\epsilon(\mathbf{r}') - 1) E(\mathbf{r}') H_0^{(1)}(k|\mathbf{r} - \mathbf{r}'|) d^2 r'$. Here, the far-field ($kr \gg 1$) scattering pattern is of a particular importance. By using the asymptotic form of the Hankel function in the far-field, one can show that for an incoming plane wave $E^{(i)} = E_0 \exp(i\mathbf{k} \cdot \mathbf{r})$, the far-field scattering response can be described via

$$E = E_0 \left( \exp(i\boldsymbol{k} \cdot \boldsymbol{r}) + f(\theta) \frac{e^{ikr}}{\sqrt{kr}} \right). \quad (3)$$

In this relation, $f(\theta)$ represents the so-called scattering amplitude, which is obtained in terms of the electric field inside the scatterer through $f(\theta) = \frac{k^2(1+i)}{2\sqrt{\pi}} \int (\epsilon(\boldsymbol{r}') - 1) E(\boldsymbol{r}') \exp(-ik\hat{\boldsymbol{r}} \cdot \boldsymbol{r}') \, d^2 r'$. Note that the scattering amplitude implicitly depends on the direction $\hat{\boldsymbol{k}}$ of the incoming plane wave.

Here, we use the method of moments as discussed in Ref. [27], in order to numerically solve the governing integral equation for the total electric field inside the scatterer. In this method, the cross section of the dielectric cylinder is divided into small cells, each radiating as an electric current filament. By assuming that the electric field is approximately constant in each cell, the integral equation is then converted into a set of $N$ linear algebraic equations [27], where $N$ represents the total number of cells. Once the total electric field inside the scatterer is known, the scattering amplitude can be subsequently obtained through the associated Fourier integral.

## III. Deflection of plane waves by PT-symmetric cylindrical objects

We now turn our attention to a PT-symmetric infinitely long dielectric cylinder, as depicted in Fig. 1. In this case, the upper half of this system displays gain, $\epsilon_1 = \epsilon_R - i\epsilon_I$, whereas the lower half a balanced amount of loss, $\epsilon_2 = \epsilon_R + i\epsilon_I$, ($\epsilon_I > 0$). The scattering strength is quantified via the following two dimensionless quantities $m_R = k_0 a (\epsilon_R - 1) = 2\pi(\epsilon_R - 1)(a/\lambda)$ and $m_I = k_0 a \epsilon_I = 2\pi\epsilon_I(a/\lambda)$. Figure 2 shows the near and far-field scattering pattern arising from such a PT- symmetric cylinder when illuminated by a plane wave along the $x$ direction. According to this figure, in the near field, light is mostly concentrated in the gain side. However, in the far field, light tends to bend toward the lossy section. Note that, aside from this deflection, the azimuthal distribution of the scattering amplitude is almost preserved. By further increasing the gain/loss contrast the bending angle increases until reaching a point where the scattering pattern changes drastically and the deflection angle cannot even be defined.

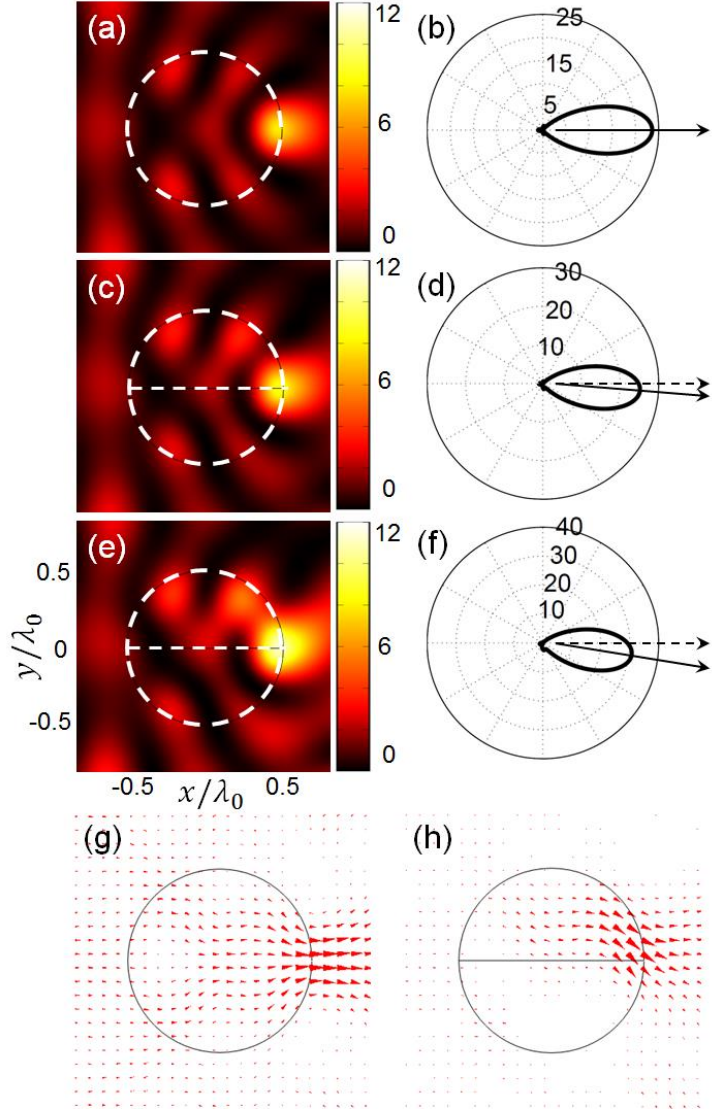

**Fig. 2.** (a,b) The near-field pattern of the total electric field intensity normalized to the incident plane wave's intensity ($|E|^2/|E_0|^2$), and far-field patterns of the scattered electric field intensity ($|f(\theta)|^2$) for the case of a passive lossless scatterer i.e., $\pm\epsilon_I = 0$, (c,d) near-field and far-field patterns for a PT-symmetric scatterer with $\pm\epsilon_I = \pm0.2$, (e,f) the same as in the previous case when the imaginary part of permittivity is increased to $\pm\epsilon_I = \pm0.4$, and (g,h) Poynting vector associated with the Hermitian and the PT-symmetric cylinders of part (a) and (e) respectively. In all cases $\epsilon_R = 2.1$, the background material is assumed to be free space ($\epsilon_b = 1$), and the diameter of the cylinder is equal to the wavelength of the incoming plane wave. In the above examples a heavy gain/loss contrast has been used to exemplify these features.

This deflection in scattering is a direct outcome of the local energy flow that takes place between the gain and the loss region. This in turn leads to a tilt in the light wavefront while propagating along the gain/loss interface of this PT-symmetric cylinder. Figures 2(g,h) depict the time average energy flux vector $S(r) = (i/4\omega\mu)(E(r)\nabla E^*(r) - E^*(r)\nabla E(r))$ in the near-field of the cylinder which clearly shows the local energy flow at the boundaries of the gain/loss regions.

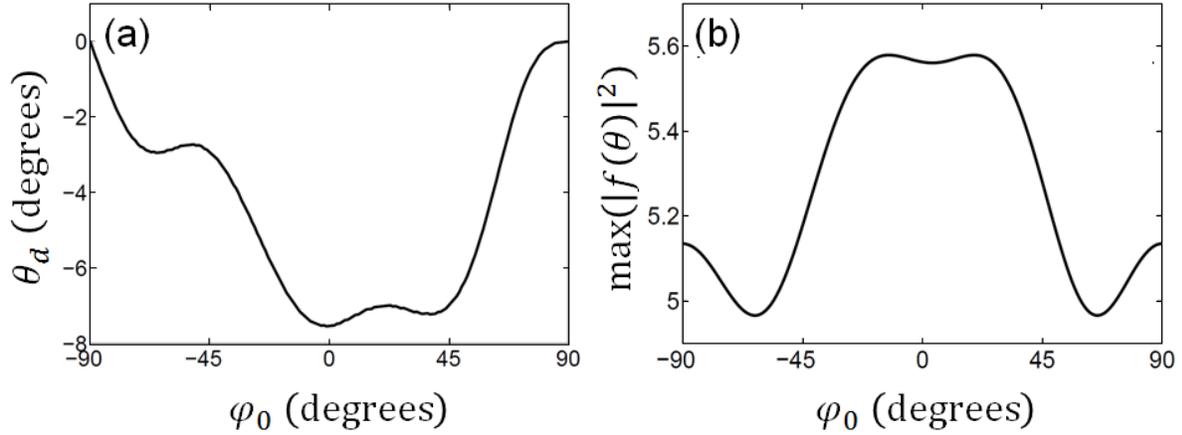

**Fig. 3.** The deflection angle (a) and the maximum scattering amplitude (b) for different angles of incidence. The PT cylinder parameters are identical to those used in Fig.2.

Following this discussion, one should expect that scattering from a PT-symmetric object must vary as a function of the incident angle with respect to the gain/loss interface. Figure 3 indeed indicates that both the angle of deflection as well as the maximum scattering amplitude change drastically with the angle of incidence. It is worth noting that, the amount of gain used in the examples of Figs. 2 and 3 is large, and therefore it is experimentally out of reach. Of interest would be to see if one can observe similar results without utilizing such high gain values. Along these lines, we consider again the PT-symmetric cylinder of the previous example while this time the gain region is replaced with a transparent material of the same relative permittivity, i.e., $\epsilon_1 = \epsilon_R$, and $\epsilon_2 = \epsilon_R + i\epsilon_I$. As shown in Fig. 4, even in the absence of gain, the aforementioned deflection property is preserved. Of course, as one should expect, the deflection angle is reduced when compared to the corresponding PT cylinder. This is because the light deflection depends on the contrast between the imaginary parts of the two regions. Note however that in absence of gain the scattering amplitude is reduced since attenuation effects cannot be compensated in the loss region.

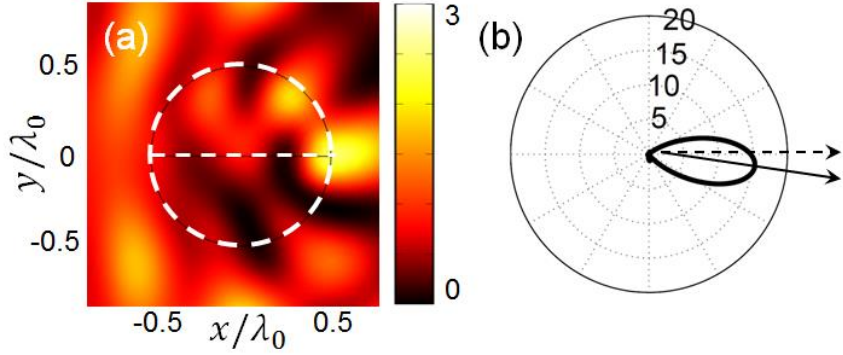

**Fig. 4.** Scattering pattern resulting from a PT-like cylinder. The real parts of the relative permittivity in both regions are the same. While the upper half of this cylinder is transparent, the other half exhibits loss. Here, $\epsilon_I = 0.4$ and all other parameters are the same as in Fig. 2.

Given that the structure of Fig. 1 lacks mirror symmetry in the $y$ direction, one may ask as to whether this deflection effect is a direct consequence of broken mirror symmetry or it is an outcome of the gain and loss present in the two sectors. To answer this question, here we consider two different scenarios. First we investigate the properties of a Janus cylinder composed of two lossless dielectric materials with different refractive indices or relative permittivities $\epsilon_1 = (\epsilon_R - \Delta\epsilon_R)$ and $\epsilon_2 = (\epsilon_R + \Delta\epsilon_R)$. In the second scenario, we consider a PT-symmetric cylinder having two parts with relative permittivities $\epsilon_1 = \epsilon_R + i\epsilon_I$ and $\epsilon_2 = \epsilon_R - i\epsilon_I$. If we assume that the index contrast in the first case ($2\Delta\epsilon_R$) is comparable to the gain and loss contrast in the second situation ($2\epsilon_I$), of interest would be to see which structure can lead to a stronger deflection. The simulation results of these two scenarios are compared in Fig. 5. As indicated in this figure, the deflection angle obtained for a PT scatterer (red) can be much larger than the one possible from a standard Janus cylinder (blue). While the difference is clearly visible for week scatterers ($m_R \ll 2\pi$), for stronger scatterers, the deflection angles corresponding to these two structures start to approach each other.

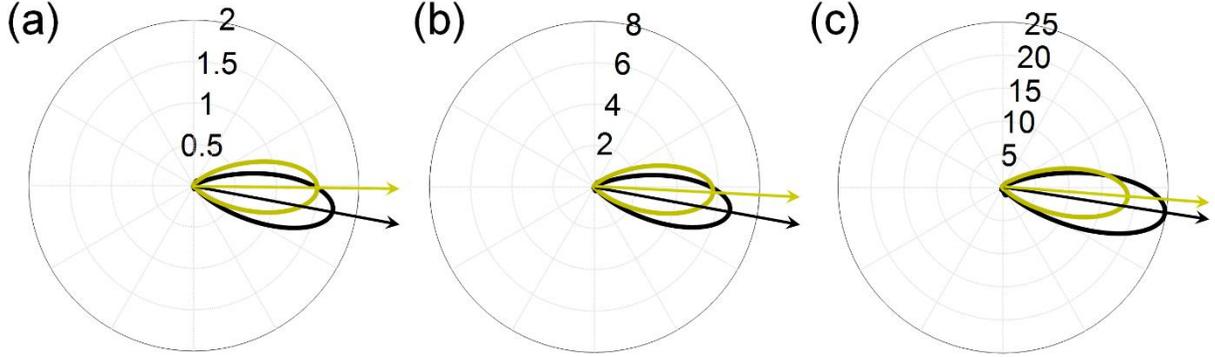

**Fig. 5.** A comparison between the far-field scattering patterns corresponding to a Janus cylinder composed of two lossless materials with different refractive indices (orange curves), and a PT-symmetric cylinder (black curves). (a) Orange: $\epsilon_{1,2} = 2.2 \mp 0.1$, black: $\epsilon_{1,2} = 2.2 \pm 0.1i$ , (b) orange: $\epsilon_{1,2} = 2.4 \mp 0.2$, black: $\epsilon_{1,2} = 2.4 \pm 0.2i$, and (c) orange: $\epsilon_{1,2} = 2.8 \mp 0.4$, black: $\epsilon_{1,2} = 2.8 \pm 0.4i$. In all cases the background material is assumed to exhibit a relative permittivity of $\epsilon_b = 2$ and the diameter of the cylinder is equal to the wavelength of the incoming light.

Before concluding this section, here we explore the effect of the size of the PT cylinder on its scattering behavior. Given that the results presented so far are all concerned with a wavelength size scatterer, of interest would be to see how the gain/loss induced bending effect is affected by different scatterer sizes. Figure 6 depicts the deflection angle $\theta_d$ with respect to the radius of the PT cylinder $a$, normalized to the free space wavelength $\lambda_0$. According to this figure, in the subwavelength regime, the deflection increases to very large angles while in the same regime the scattering amplitude is weak and spans the entire range of the polar angle $\theta$. In the opposite regime, however, when the scatterer's size is larger than the wavelength, the far-field scattering pattern becomes very narrow. In this case, the scattered light is still deflected with respect to the incoming plane wave but the deflection angle is much smaller.

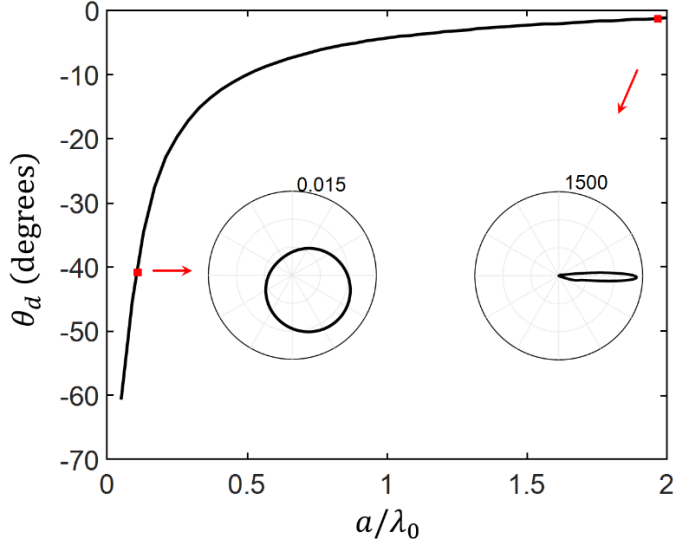

**Fig. 6.** The deflection angle $\theta_d$ versus the normalized radius $a/\lambda_0$ of a PT-symmetric cylinder with $\epsilon_{1,2} = 2.4 \pm i\, 0.2$ in a background material with $\epsilon_b = 2$. The two insets show the far-field scattering pattern at two specific points shown with red dots.

## IV. Modified optical theorem for PT-symmetric scatterers

According to Figs. 2 (a,b), for the fully transparent case, maximum scattering amplitude occurs right behind the cylinder. Indeed, the optical theorem demands that the scattering amplitude right behind a Hermitian scatterer is never zero. The optical theorem itself is an outcome of the power conservation in Hermitian systems. In such arrangements this theorem relates the total scattered power to the scattering amplitude right behind the scatterer. For 2D structures and under TE polarization conditions, this relation can be stated as $\int_0^{2\pi} |f(\theta)|^2 d\theta = -2\sqrt{\pi}\Re[(1+i)f(0)]$ where $\Re$ represents the real part [28].

Even though in the presence of gain and loss power conservation does not necessarily hold in the usual way, because the gain region can supply additional energy, as we show here, the optical theorem can be accordingly modified for PT-symmetric structures. To show this, let us start with Eq. (2) in the presence of a PT-symmetric relative permittivity distribution, as given by Eq. (1). Using Eq. (2) along with its parity and time reversed counterpart one can simply show that

$$E(r)\nabla^2 E^*(-r) - E^*(-r)\nabla^2 E(r) = 0. \quad (4)$$

By integrating this relation over the area of a circle of radius $r \to \infty$ which spans over the entire $xy$ plane, we find $\int_0^{2\pi}(E(r)\cdot \nabla E^*(-r) - E^*(-r)\cdot \nabla E(r))\cdot \hat{r}d\theta = 0$. Without any loss of generality, we assume that the incoming wave is propagating along the $x$ direction $\mathbf{k} = \hat{x}k$, and hence, by using the far-field approximation (Eq. (3)), the electric fields can be written as $E(r) = e^{ikr\cos\theta} + f(\theta)\frac{e^{ikr}}{\sqrt{kr}}$ and $E^*(-r) = e^{ikr\cos\theta} + f^*(\theta + \pi)\frac{e^{-ikr}}{\sqrt{kr}}$. By employing these approximations in the integral relation, and after neglecting terms that decay faster than $(kr)^{-1}$, one finds that $\frac{2}{kr}\int_0^{2\pi} f(\theta)f^*(\theta + \pi)d\theta + \frac{1}{\sqrt{kr}}\int_0^{2\pi}(1 + \cos\theta)f^*(\theta + \pi)e^{-ikr(1-\cos\theta)}d\theta + \frac{1}{\sqrt{kr}}\int_0^{2\pi}(1 - \cos\theta)f(\theta)e^{ikr(1+\cos\theta)}d\theta = 0$. The stationary phase approximation can now be used to show that:

$$\int_0^{2\pi} f(\theta)f^*(\theta + \pi)d\theta = -2\sqrt{\pi}\mathfrak{R}[(1 + i)f(\pi)], \quad (5)$$

which is a modified version of the optical theorem valid for PT-symmetric objects. Equation (5) can be viewed as a manifestation of the quasi-energy conservation in PT-symmetric systems. Notice that, differently from the conventional optical theorem, this sum rule relates the integral on the left to the backward scattering from the PT-symmetric object. The quantity in the integrand can be in general complex, implying that zero backscattering is not necessarily synonymous of zero scattering at all angles, but only that the integrated quantity is zero.

To further explain the meaning of Eq. (5), we define a time averaged quasi-energy flux vector as:

$$\mathbf{F}(r) = \frac{i}{4\omega\mu}[E(r)\nabla E^*(-r) - E^*(-r)\nabla E(r)]. \quad (6)$$

Note that Eq. (4) directly follows $\nabla \cdot \mathbf{F} = 0$, thus the line integral of $\mathbf{F}$ over an arbitrary closed contour is zero:

$$\oint \mathbf{F}\cdot \hat{n}d\ell = 0, \quad (7)$$

where $\hat{\boldsymbol{n}}$ represents the local normal vector to the contour. It should be noted that this latter relation shows the conservation of the quasi-energy, and it is derived directly from Eqs. (1,2), thus it is valid for any PT-symmetric structure. A specific integration contour in Eq. (7) would be a circle of radius $r$ centering on the origin that in the asymptotic limit of $r \to \infty$ leads to the PT optical theorem (Eq. (5)). Therefore Eq. (5) can be described as a relation between the rate of quasi-energy scattering by a PT-symmetric structure and its field scattering amplitude in the backward direction. Another interesting outcome of this relation is that the far-field integral of $\int_0^{2\pi} f(\theta) f^*(\theta + \pi) d\theta$ for a PT-symmetric structure is a real quantity which can be either positive or negative.

## V. Conclusion

In conclusion, we have studied the scattering properties of PT-symmetric cylinders. We showed that such scatterers can deflect light toward the loss sector at an angle that depends on the gain/loss contrast. Such bending effect depends also on the angle of incidence in a way that maximum deflection is obtained when the incoming wave propagates along the gain and loss interface. We showed that similar results can be obtained for quasi-PT structures which involve only loss. The effect of the scatterer size was explored and it was shown that the bending effect is stronger in the subwavelength regime where the scattering strength itself is weak. Finally, we investigated the optical theorem associated with two-dimensional PT-symmetric structures. We derived a modified optical theorem for PT scatterers which reflects the underlying symmetries of PT-symmetric structures and is a manifestation of quasi-energy conservation in such systems. Our results can be potentially useful for applications where beam deflection over a wavelength scale device is required.